\documentclass[12pt]{article}

\usepackage{scicite}

\usepackage{times}

\usepackage{graphicx,psfrag}
\usepackage{epstopdf}
\usepackage{amsmath}
\usepackage{bm}
\usepackage{amssymb}
\usepackage{subfigure}
\usepackage{color}
\usepackage{ulem}

\newenvironment{sciabstract}{%
\begin{quote} \bf}
{\end{quote}}

\title{Thermoacoustics of solids: a pathway to solid state engines and refrigerators}

\author
{Haitian Hao,$^{1}$ Carlo Scalo,$^{1}$ Mihir Sen,$^{2}$ Fabio Semperlotti$^{1\ast}$ \\
\\
\normalsize{$^{1}$School of Mechanical Engineering, Purdue University, West Lafayette, IN 47907, USA}\\
\normalsize{$^{2}$Aerospace and Mechanical Engineering, University of Notre Dame, Notre Dame, IN 46556, USA}\\
\\
\normalsize{$^\ast$To whom correspondence should be addressed; E-mail:  fsemperl@purdue.edu.}
}

\date{}

\begin{document} 


\baselineskip24pt


\maketitle


\begin{sciabstract}
 Thermoacoustic oscillations have been one of the most exciting discoveries of the physics of fluids in the 19th century. Since its inception, scientists have formulated a comprehensive theoretical explanation of the basic phenomenon which has later found several practical applications to engineering devices. To-date, all studies have concentrated on the thermoacoustics of fluid media where this fascinating mechanism was exclusively believed to exist. Our study shows theoretical and numerical evidence of the existence of thermoacoustic instabilities in solid media. Although the underlying physical mechanism is analogous to its counterpart in fluids, the theoretical framework highlights relevant differences that have important implications on the ability to trigger and sustain the thermoacoustic response. This mechanism could pave the way to the development of highly robust and reliable solid-state thermoacoustic engines and refrigerators.
\end{sciabstract}

\paragraph*{One sentence Summary} This paper provides the first theoretical study and numerical validation showing the existence of heat-induced, self-amplifying thermoacoustic oscillations in solids.

\paragraph*{Introduction.}
The existence of thermoacoustic oscillations in thermally-driven fluids and gases has been known for centuries. When a pressure wave travels in a confined gas-filled cavity while being provided heat, the amplitude of the pressure oscillations can grow unbounded. This is a self-sustaining process that builds upon the dynamic instabilities that are intrinsic in the thermoacoustic process.
In 1850, Soundhauss \cite{Soundhauss} experimentally showed the existence of heat-generated sound during a glassblowing process. Few years later (1859), Rijke \cite{Rijke} discovered another method to convert heat into sound based on a heated wire gauze placed inside a vertically-oriented open tube. He observed self-amplifying vibrations that were maximized when the wire gauze was located at one fourth the length of the tube. Later, Rayleigh \cite{Rayleigh} presented a theory able to qualitatively explain both Soundhauss and Rijke thermoacoustic oscillations phenomena. In 1949, Kramers \cite{Kramers} was the first to start the formal theoretical study of thermoacoustics by extending Kirchhoff's theory of the decay of sound waves at constant temperature \cite{Kirchhoff} to the case of attenuation in presence of a temperature gradient. Rott et al. \cite{Rott1,Rott2,Rott3,Rott4,Rott5,Rott6,Rott7,Rott8} made key contributions to the theory of thermoacoustics by developing a fully analytical, quasi-one-dimensional, linear theory that provided excellent predictive capabilities. It was mostly Swift \cite{Swift}, at the end of the last century, who started a prolific series of studies dedicated to the design of various types of thermoacoustic engines based on Rott's theory. Since the development of the fundamental theory, many studies have explored practical applications \cite{Jin, Chen, Garrett} of the thermoacoustic phenomenon with particular attention to the design of engines and refrigerators. 

To-date, thermoacoustic instabilities have been theorized and demonstrated only for fluids. In this study, we provide theoretical and numerical evidence of the existence of this phenomenon in solid media. In particular, we show that a solid metal rod subject to a prescribed temperature gradient on its outer boundary can undergo self-sustained vibrations driven by a thermoacoustic instability phenomenon. 

In the following, we first introduce the theoretical framework that uncovers the existence and the fundamental mechanism at the basis of the thermoacoustic instability in solids. Then, we provide numerical evidence to show that the instability can be effectively triggered and sustained. We anticipate that, although the fundamental physical mechanism resembles the thermoacoustic of fluids, the different nature of sound and heat propagation in solids produces noticeable differences in the theoretical formulations and in the practical implementations of the phenomenon. 

\begin{figure}[h]
	\centering
	\includegraphics[scale=0.5]{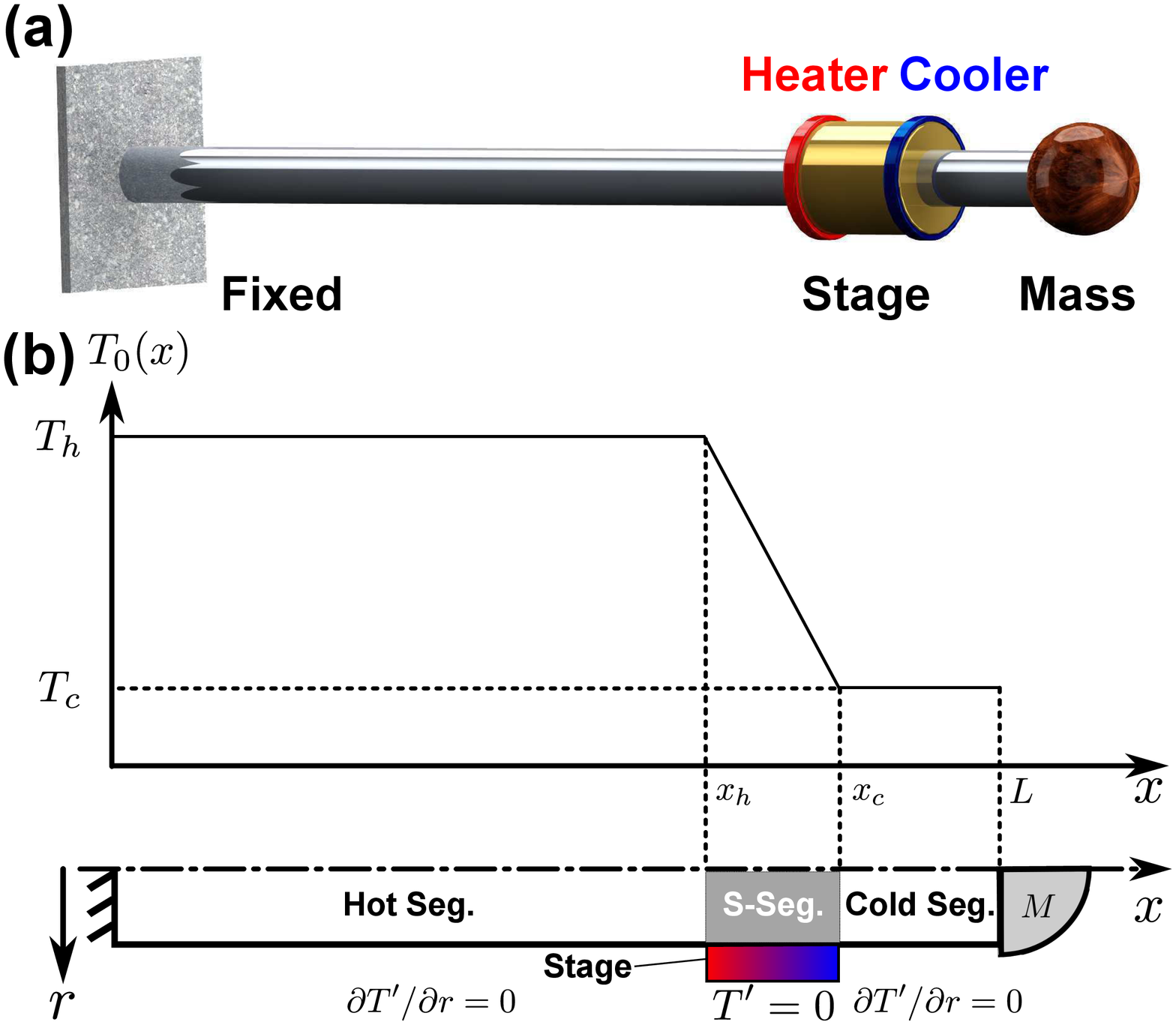}
	\caption{ (\textbf{a}) Notional schematic of the system exhibiting thermoacoustic response. An aluminum rod with circular cross-section under fixed-free boundary conditions. The free end carries a concentrated mass used to tune the frequency of the longitudinal resonance and the corresponding wavelength. A section of the rod is surrounded by a large thermal inertia (represented by a highly-thermally-conductive solid) on which a heater and a cooler are connected in order to create a predefined thermal gradient; this component is referred to as a \textit{stage}. The stage is the equivalent of the \textit{stack} in classical thermoacoustic setups. The rod is free to slide in the stage (i.e. negligible mechanical shear forces) while the heat transfer normal to the rod-stage surface is guaranteed by highly-thermally conductive grease. Heat insulating material (not showed) is assumed to be placed around the rod to reduce radiative heat losses and therefore approximate adiabatic boundary conditions. (\textbf{b}) (top) idealized reference temperature profile $T_0(x)$ produced along the rod, and (bottom) schematic of an axi-symmetric cross section of the rod showing the characteristic geometric parameters and the correspondence to the temperature profile. Three relevant segments are identified: \textit{1) S-segment}, \textit{2) hot segment}, \textit{3) cold segment}. These three segments correspond to the isothermal and the two adiabatic boundary conditions, respectively.}
	\label{rodT0}
\end{figure}

\paragraph*{Problem statement.} The fundamental system under investigation consists of a slender solid metal rod with circular cross section (Fig. \ref{rodT0}). The rod is subject to a temperature (spatial) gradient applied on its outer surface at a prescribed location, while the remaining sections have adiabatic boundary conditions. We investigate the coupled thermoacoustic response that ensues as a result of an externally applied thermal gradient and of an initial mechanical perturbation of the rod.

We anticipate that the fundamental dynamic response of the rod is governed by the laws of thermoelasticity. According to classical thermoelasticity \cite{Nowacki,Eringen,Chadwick}, an elastic wave traveling through a solid medium is accompanied by a thermal wave, and viceversa. The thermal wave follows from the thermoelastic coupling which produces local temperature fluctuations (around an average constant temperature $T_0$) as a result of a propagating stress wave.
When the elastic wave is not actively sustained by an external mechanical source, it attenuates and disappears over a few wavelengths due to the presence of dissipative mechanisms (such as, material damping); in this case the system has a positive decay rate (or, equivalently, a negative growth rate). In the ideal case of an undamped thermoelastic system, the mechanical wave does not attenuate but, nevertheless, it maintains bounded amplitude. In such situation, the total energy of the system is conserved (energy is continuously exchanged between the thermal and mechanical waves) and the stress wave exhibits a zero decay rate (or, equivalently, a zero growth rate).

Contrarily to the classical thermoelastic problem where the medium is at a uniform reference temperature $T_0$ with an adiabatic outer boundary, when the rod is subject to heat transfer through its boundary (i.e. non-adiabatic conditions) the thermoelastic response can become unstable. In particular, when a proper temperature spatial gradient is enforced on the outer boundary of the rod then the initial mechanical perturbation can grow unbounded due to the coupling between the mechanical and the thermal response. This last case is the exact counterpart that leads to thermoacoustic response in fluids, and it is the specific condition analyzed in this study. For the sake of clarity, we will refer to this case, which admits unstable solutions, as the thermoacoustic response of the solid (in order to differentiate it from the classical thermoelastic response). 

\paragraph*{Thermoacoustic model for solids.} In order to show the existence of the thermoacoustic phenomenon in solids, we developed a theoretical three-dimensional model describing the fully-coupled thermoacoustic response. The model builds upon the classical thermoelastic theory developed by Biot \cite{Biot} further extended in order to account for coupling terms that are key to capture the thermoacoustic instability. Starting from the fundamental conservation principles (see \cite{Suppl} for the detailed analytical derivation), the nonlinear thermoacoustic equations for a homogeneous isotropic solid in an Eulerian reference frame are written as:

\begin{align}
\rho \frac{Dv_i}{Dt}&=\sum_{j=1}^{3}\frac{\partial\sigma_{ji}}{\partial x_j}+F_{b,i} \label{mom}\\
\rho c_\epsilon \frac{DT}{Dt}+\frac{\alpha E T}{1-2\nu} \frac{D e_v}{D t} &= \sum_{j=1}^{3}\frac{\partial}{\partial x_j}\bigg({\kappa \frac{\partial T}{\partial x_j}}\bigg)+\dot{q_g} \qquad\qquad (i,j=1,2,3) \label{ene}
\end{align}
where Eqns. \ref{mom} and \ref{ene} are the conservation of momentum and energy, respectively. In the above equations $\rho$ is the material density, $E$ is the Young's modulus, $\nu$ is the Poisson's ratio, $\alpha$ is the thermoelastic expansion coefficient, $c_\epsilon$ is the specific heat at constant strain, $\kappa$ is the thermal conductivity of the medium, $v_i$ is the particle velocity in the $x_i$ direction, $\sigma_{ji}$ is the stress tensor, $D/Dt=\frac{\partial()}{\partial t}+\Sigma_{i=1}^3 v_i \cdot \frac{\partial()}{\partial x_i}$ is the material derivative, $T$ is the total temperature, and $e_v$ is the volumetric dilatation which is defined as $e_v=\sum_{j=1}^{3} \varepsilon_{jj}$. $F_{b,i}$ and $\dot{q_g}$ are the mechanical and thermal source terms, respectively. The stress-strain constitutive relation for a linear isotropic solid, including the Duhamel components of temperature induced strains, is given by:
\begin{align}
\sigma_{ij}=2\mu \varepsilon_{ij}+\bigg[\lambda_L e_v - \alpha(2\mu+3\lambda_L) (T-T_0)\bigg]\delta_{ij} \label{con}
\end{align}
where $\mu$ and $\lambda_L$ are the Lam\'{e} constants, $\varepsilon_{ij}$ is the strain tensor,  $T_0$ is the mean temperature, and $\delta_{ij}$ is the Kronecker delta.

The fundamental element for the onset of the thermoacoustic instability is the application of a thermal gradient. In classical thermoacoustics of fluids, the gradient is applied by using a stack element which enforces a linear temperature gradient over a selected portion of the domain. The remaining sections are kept under adiabatic conditions. In analogy to this traditional thermoacoustic design, we enforced the thermal gradient using a \textit{stage} element that can be thought as the equivalent of a single-channel stack. Upon application of the stage, the rod could be virtually divided in three segments: the hot segment, the S-segment, and the cold segment (Fig. \ref{rodT0}b). The hot and cold segments were kept under adiabatic boundary conditions. The S-segment was the region underneath the stage, where the spatial temperature gradient was applied and heat exchange could take place.

Under the conditions described above, the governing equations can be solved in order to show that the dynamic response of the solid accepts thermoacoustically unstable solutions. In the following, we use a two-fold strategy to characterize the response of the system based on the governing equations (Eqns. \ref{mom} and \ref{ene}). First, we linearize the governing equations and synthesize a quasi-one-dimensional theory in order to carry on a stability analysis. This approach allows us to get deep insight into the material and geometric parameters contributing to the instability. Then, in order to confirm the results from the linear stability analysis and to evaluate the effect of the nonlinear terms, we solve numerically the 3D nonlinear model to evaluate the response in the time domain.

Before concluding this section we should point out a noticeable difference of our model with respect to the classical thermoelastic theory of solids. Due to the existence of a mean temperature gradient $T_0(x)$, the convective component of the temperature material derivative is still present, after linearization, in the energy equation. This term typically cancels out in classical thermoelasticity, given the traditional assumption of a uniform background temperature $T_0=const.$, while it is the main driver for thermally-induced oscillations.

\paragraph*{Quasi-1D theory: linear stability analysis.}

In order to perform a stability analysis, we first extract the one-dimensional governing equations from Eqns. \ref{mom} and \ref{ene} and then proceed to their linearization. The linearization is performed around the mean temperature $T_0(x)$, which is a function of the axial coordinate $x$. The mean temperature distribution in the hot segment $T_h$ and in the cold segment $T_c$ are assumed constant. Note that even if these temperature profiles were not constant, the effect on the instability would be minor as far as the segments were maintained in adiabatic conditions \cite{Suppl}. The $T_0$ profile on the isothermal section follows from a linear interpolation between $T_h$ and $T_c$ (see Fig. \ref{rodT0}).

The following quasi-1D analysis can be seen as an extension to solids of the well-known Rott's stability theory. We use the following assumptions: a) the rod is axisymmetric, b) the temperature fluctuations caused by the radial deformation are negligible, and c) the axial thermal conduction of the rod is also negligible (the implications of this last assumption are further discussed in \cite{Suppl}).
According to Rott's theory, we transform Eqns. \ref{mom} and \ref{ene} to the frequency domain under the \textit{Ansatz} that all fluctuating (primed) variables are harmonic in time. This is equivalent to $()^\prime=()-()_0=\hat{()}e^{i\Lambda t}$, where $\hat{()}$ is regarded as the fluctuating variable in frequency domain. $\Lambda=-i\beta+\omega$, $\omega$ is the angular frequency of the harmonic response, and $\beta$ is the growth rate (or the decay rate, depending on its sign). By substituting Eqn. \ref{con} in Eqn. \ref{mom} and neglecting the source terms, the set of linearized quasi-1D equations \cite{Scalo} are: 
\begin{align}
i\Lambda \hat{u}&=\hat{v}\\
i\Lambda \hat{v}&=\frac{E}{\rho}\bigg(\frac{d^2\hat{u}}{dx^2}-\alpha\frac{d\hat{T}}{dx}\bigg) \label{1dmom}\\
i\Lambda \hat{T}&=-\frac{dT_0}{dx}\hat{v}-\gamma_G T_0\frac{d\hat{v}}{dx}-\alpha_H\hat{T} \label{1dene}
\end{align}
where $\gamma_G=\frac{\alpha E}{\rho c_\epsilon (1-2\nu)}$ is the $Gr\ddot{u}neisen$ constant \cite{Yates},  $i$ is the imaginary unit, $\hat{u}$, $\hat{v}$ and $\hat{T}$ are the fluctuations of the particle displacement, particle velocity, and temperature averaged over the cross section of the rod. For brevity, they will be referred to as fluctuation terms in the following. The intermediate transformation $i\Lambda \hat{u}=\hat{v}$ avoids the use of quadratic terms in $\Lambda$, which ultimately enables the system to be fully linear. The $\alpha_H \hat{T}$ term in Eqn. \ref{1dene} accounts for the thermal conduction in the radial direction, and it is the term that renders the theory quasi-1D. The function $\alpha_H$ is given by: 

\begin{align}
\alpha_H=
\begin{cases}
\dfrac{\omega \xi_{top} \dfrac{J_1(\xi_{top})}{J_0(\xi_{top})}}{i \xi_{top} \dfrac{J_1(\xi_{top})}{J_0(\xi_{top})} - \dfrac{R^2}{\delta_k^2}} &  x_h<x<x_c\\
0 &\text{elsewhere}
\end{cases}
\end{align}
where $J_n(\cdot)$ are Bessel functions of the first kind, and $\xi$ is a dimensionless complex radial coordinate given by:
\begin{align}
\xi=\sqrt{-2i}\frac{r}{\delta_k}
\end{align}
thus, the dimensionless complex radius is
$\xi_{top}=\sqrt{-2i}\frac{R}{\delta_k}$,
where $R$ is the radius of the rod. The thermal penetration thickness $\delta_k$ is defined as $\delta_k=\sqrt{\frac{2\kappa}{\omega \rho c_\epsilon}}$, and physically represents the depth along the radial direction (measured from the isothermal boundary) that heat diffuses through. The full derivation of the one-dimensional equations can be found in \cite{Suppl}.

The one-dimensional model was used to perform a stability eigenvalue analysis. The eigenvalue problem is given by $(i\Lambda {\bf I} - {\bf A} ){\bf y} = {\bf 0}$ where ${\bf I}$ is the identity matrix, ${\bf A}$ is a matrix of coefficients, $\textbf{0}$ is the null vector, and ${\bf y}=[{\bf\hat{u}};{\bf\hat{v}};{\bf\hat{T}}]$ is the vector of state variables where ${\bf\hat{u}},{\bf\hat{v}}$, and ${\bf\hat{T}}$ are the particle displacement, particle velocity, and temperature fluctuation eigenfunctions.

The eigenvalue problem was solved numerically for the case of an aluminum rod having a length of $L=1.8m$ and a radius $R=2.38mm$. The following material parameters were used: density $\rho=2700 kg/m^3$, Young's modulus $E=70 GPa$, thermal conductivity $\kappa=238 W/(m\cdot K)$, specific heat at constant strain $c_\epsilon=900 J/(kg\cdot K)$, and thermal expansion coefficient $\alpha=23\times 10^{(-6)} K^{-1}$. The strength of the instability in classical thermoacoustics (often quantified in terms of the ratio $\beta/\omega$) depends, among the many parameters, on the location of the thermal gradient. This location is also function of the wavelength of the acoustic mode that triggers the instability, and therefore of the specific (mechanical) boundary conditions. We studied two different cases: 1) \textit{fixed-free} and 2) \textit{fixed-mass}. In the fixed-free boundary condition case, the optimal location of the stage was approximately around $1/2$ of the total length of the rod, which is consistent with the design guidelines from classical thermoacoustics. Considerations on the optimal design and location of the stage/stack will be addressed in the following section; at this point we assumed a stage located at $x=0.5L$ with a total length of $0.05L$. 

Assuming a mean temperature profile equal to $T_h=493.15K$ in the hot part and to $T_c=293.15K$ in the cold part, the 1D theory returned the fundamental eigenvalue to be $i\Lambda=0.404+i4478(rad/s)$. The existence of a positive real component of the eigenvalue revealed that the system was unstable and self-amplifying, that is it could undergo growing oscillations as a result of the positive growth rate $\beta$. The growth ratio was found to be $\beta/\omega=9.0\times10{-5}$.

Equivalently, we analyzed the second case with fixed-mass boundary conditions. In this case, a $2kg$ tip mass was attached to the free end with the intent of tuning the resonance frequency of the rod and increasing the growth ratio $\beta/\omega$ which controls the rate of amplification of the system oscillations. An additional advantage of this configuration is that the operating wavelength increases. To analyze this specific boundary condition configuration, we chose $x_h=0.9L$ and $x_c-x_h=0.05L$.
The stability analysis returned the first eigenvalue as $i\Lambda=0.210+i585.5(rad/s)i$ resulting in a growth ratio $\beta/\omega=3.6\times10^{-4}$, larger than the \textit{fixed-free} case.

The above results from the quasi-1D thermoacoustic theory provided a first important conclusion of this study, that is confirming the existence of thermoacoustic instabilities in solids as well as their conceptual affinity with the analogous phenomenon in fluids.

\begin{figure}[h]
	\centering
	\includegraphics[scale=0.5]{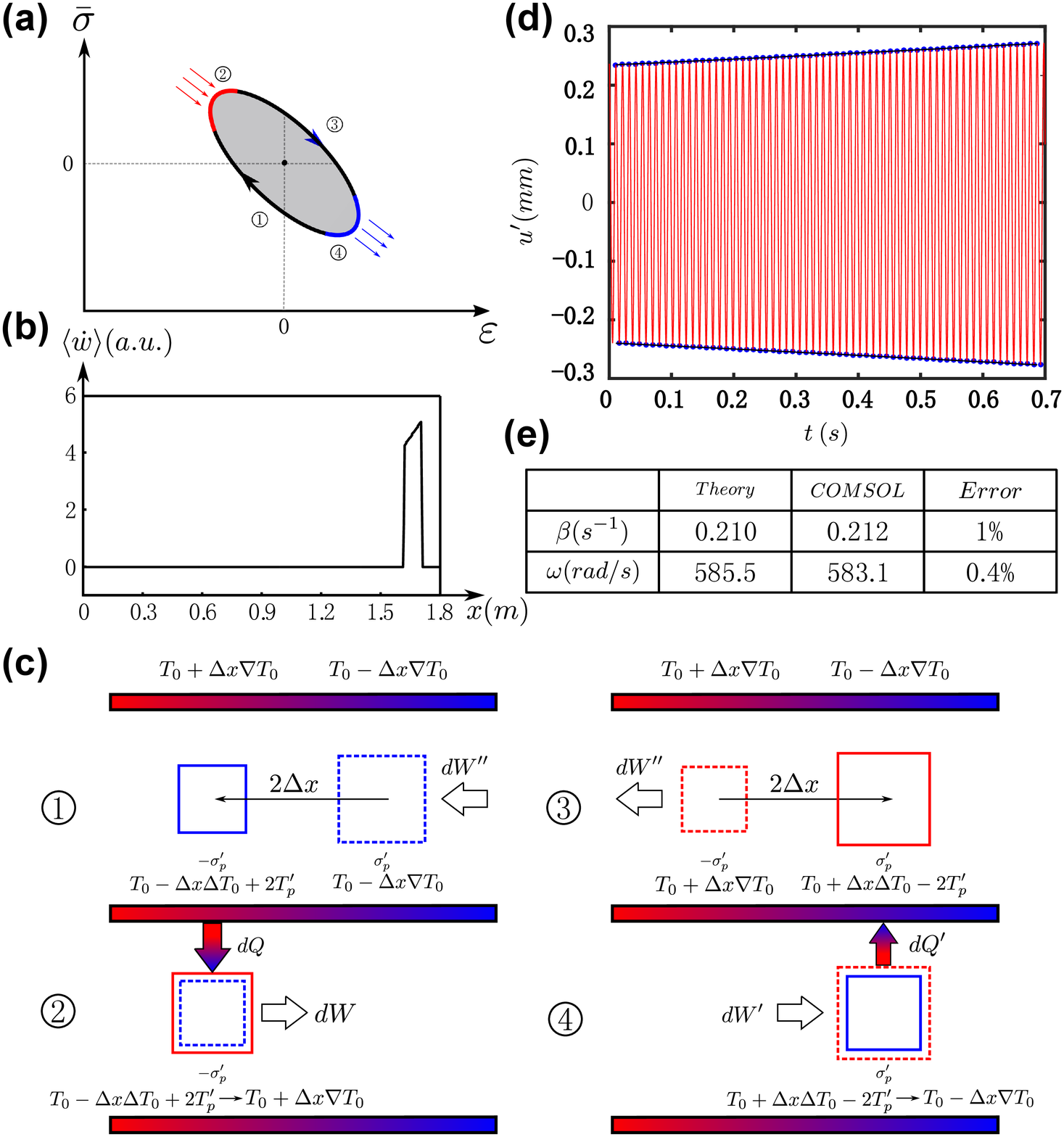}
	\caption{(\textbf{a}) Schematic of the thermodynamic cycle of a Lagrangian particle in the S-segment  during an acoustic/elastic cycle (see also \cite{Suppl}). (\textbf{b}) The time averaged volume-change work $\langle{\dot{w}}\rangle$ (presented in arbitrary scale and units) along the length of the rod showing that the net work is generated in the stage. (\textbf{c}) Schematic view showing the evolution of an infinitesimal volume element during the different phases of the thermodynamic cycle (a). For simplicity, the cycle is divided in two reversible adiabatic steps and two irreversible constant-stress steps. $()_p^\prime$ indicates the peak value of the corresponding fluctuating variables. (\textbf{d}) Time history of the axial displacement fluctuation at the end of the rod for the fixed-mass configuration. `\textit{Red} \textendash': Response, `\textit{Blue} \textbullet': Peak values, `\textit{Black} \textendash': Exponential fit. (\textbf{e}) Table presenting a comparison of the results between the quasi-1D theory and the numerical FE 3D model.  }
	\label{comsol}
\end{figure}

To get a deeper physical insight into this phenomenon, we studied the themodynamic cycle of a particle located in the S-region. The mechanical work transfer rate or, equivalently, the volume-change work per unit volume may be defined as $\dot{w}=-\sigma\frac{\partial \varepsilon}{\partial t}$ \cite{Barron}, where $\sigma$ and $\varepsilon$ are the total axial stress (i.e. including both mechanical and thermal components) and strain, respectively. During one acoustic/elastic cycle, the time averaged work transfer rate per unit volume is $\langle{\dot{w}}\rangle=\frac{1}{\tau}\int_0^\tau (-\sigma)\frac{\partial \varepsilon}{\partial t} dt= \frac{1}{\tau}\int_0^\tau (-\sigma) d\varepsilon=\frac{1}{\tau}\int_0^\tau \bar\sigma d\varepsilon $, where $\tau$ is the period of a cycle, and $\bar \sigma=(-\sigma)$. Figure \ref{comsol}a shows the $\bar\sigma$-$\varepsilon$ diagram where the area enclosed in the curve represents the work per unit volume done by the infinitesimal volume element in one cycle. All the particles located in the regions outside the S-segment do not do net work because the temperature fluctuation $T^\prime$ is in phase with the strain $\varepsilon$, which ultimately keeps the stress and strain in phase (thus, the area enclosed is zero). Figure \ref{comsol}b shows the time-averaged work $\langle{\dot{w}}\rangle=\frac{1}{2}Re[\hat{\bar{\sigma}}(i\omega\hat{\varepsilon})^*]$ along the rod, where $()^*$ denotes the complex conjugate. Note that the rate of work $\langle{\dot{w}}\rangle$ was evaluated based on modal stresses and strains, therefore its value must be interpreted on an arbitrary scale. The large increase of $\langle{\dot{w}}\rangle$ at the stage location indicates that a non-zero net work is only done in the section where the temperature gradient is applied (and therefore where heat transfer through the boundary takes place).
		
Figure \ref{comsol}c shows a schematic representation of the thermo-mechanical process taking place over an entire vibration cycle. When the infinitesimal volume element is compressed, it is displaced along the $x$ direction while its temperature increases (step 1). As the element reaches a new location, heat transfer takes place between the element and its environment. Assuming that in this new position the element temperature is lower than the surrounding temperature, then the environment provides heat to the element causing its expansion. In this case, the element does net work $dW$ (step 2) due to volume change. Similarly, when the element expands (step 3), the process repeats analogously with the element moving backwards towards the opposite extreme where it encounters surrounding areas at lower temperature so that heat is now extracted from the particle (and provided to the stage). In this case, work $dW^\prime$ is done on the element due to its contraction (step 4). The net work generated during one cycle is $dW-dW^\prime$.

\paragraph*{3D model: time-dependent analysis.}
In order to validate the quasi-1D theory and to estimate the possible impact of three-dimensional and nonlinear effects, we solved the full set of Eqns. \ref{mom} and \ref{ene} in the time domain. The equations were solved by finite element method on a three-dimensional geometry using the commercial software \textit{Comsol Multiphysics}. We highlight that with respect to Eqns. \ref{mom} we drop the nonlinear convective derivative $v_i \frac{\partial v_i}{\partial x_i}$ which effectively results in the linearization of the momentum equation. Full nonlinear terms are instead retained in the energy equation. The details of the analysis are provided in \cite{Suppl}.

Figure \ref{comsol}d shows the time history of the axial displacement fluctuation $u'$ at the free end of the rod. The dominant frequency of the oscillation is found, by Fourier transform, to be equal to $\omega=583.1(rad/s)$, which is within $0.4\%$ from the prediction of the 1D theory. The time response is evidently growing in time therefore showing clear signs of instability. The growth rate was estimated by either a logarithmic increment approach or an exponential fit on the envelope of the response. The logarithmic increment approach returns $\beta$ as:
\begin{align}
\beta=\frac{1}{N-1}\sum_{i=2}^{N} ln\frac{A_i}{A_1}/(t_i-t_1)
\end{align}
where $A_1$ and $A_i$ are the amplitudes of the response at the time instant $t_1$ and $t_i$, and where $t_1$ and $t_i$ are the start time and the time after $(i-1)$ periods. Both approaches return $\beta=0.212(rad/s)$. This value is found to be within $1\%$ accuracy from the value obtained via the quasi-1D stability analysis, therefore confirming the validity of the 1D theory and of the corresponding simplifying assumptions.

\paragraph*{Discussion.}
In reviewing the thermoacoustic phenomenon in both solids and fluids we note similarities as well as important differences between the underlying mechanisms. These differences are mostly rooted in the form of the constitutive relations of the two media. 
Both the longitudinal mode and the transverse heat transfer are pivotal quantities in thermal-induced oscillations of either fluids or solids. The longitudinal mode sustains the stable vibration and provides the necessary energy flow, while the transverse heat transfer controls the heat and momentum exchange between the medium and the stage/stack.

The growth rate of the mechanical oscillations is affected by several parameters including the amplitude of the temperature gradient, the location of the stage, the thermal penetration thickness, and the energy dissipation in the system. Here below, we investigate these elements individually. The effect of the temperature gradient is straightforward because higher gradients result in higher growth rate.

\begin{figure}
	\includegraphics[scale=0.5]{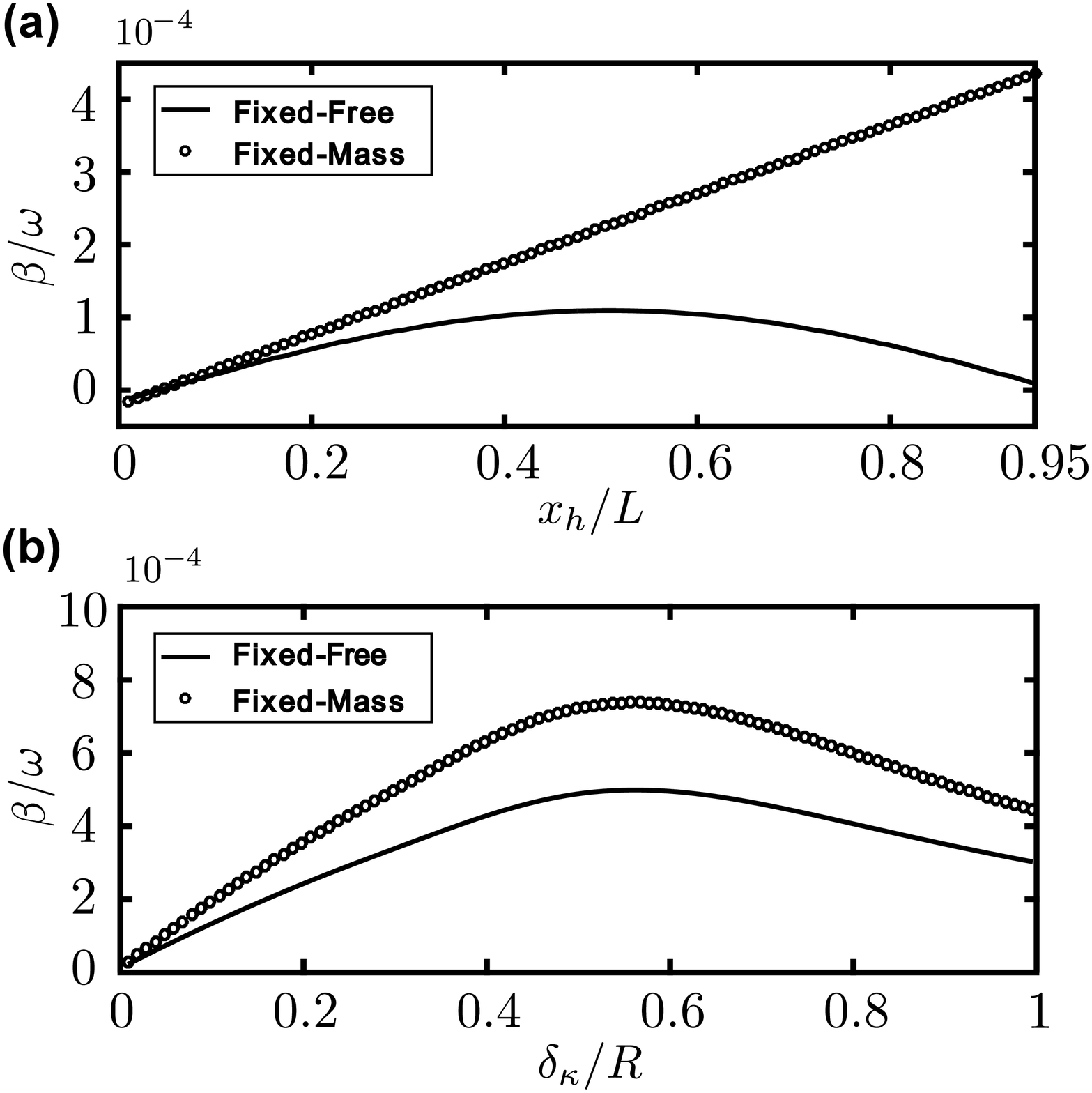}
	\caption{Plots of the growth ratio versus different non-dimensional parameters showing the existence of optimal values. {\bf (a)}. The growth ratio versus the location of the stage non-dimensionalized by the length $L$ of the rod. Results for a stage length $\Delta L=5\%L$ and radius $R$ of the rod fixed. {\bf (b)}. The growth ratio versus the penetration thickness non-dimensionalized by the rod radius $R$. Results obtained for a fixed location and length of the stage. }
	\label{opt}
\end{figure}

The location of the stage relates to the phase lag between the particle velocity and the temperature fluctuations, which is one of the main driver to achieve the instability. In fluids, the optimal location of the stack in a tube with closed ends is about one-forth the tube length, measured from the hot end. In a solid, we show that the optimal location of the stage is at the midspan for the fixed-free boundary condition, and at the mass end for the fixed-mass boundary condition (Fig. \ref{opt}a). This conclusion is consistent with similar observations drawn in thermoacoustics of fluids where a closed tube (equivalent to a fixed-fixed boundary condition in solids) gives a half-wavelength tube ($L_{0.5}=\frac{1}{2}\lambda$, where $\lambda$ indicates wavelength, $L_{0.5}$ and $L_{0.25}$ length of a half- and quater-wavelength rod/tube respectively). The optimal location, 1/4 tube length, is equivalent to 1/8 wavelength ($x_{opt}=\frac{1}{4}L_{0.5}=\frac{1}{8}\lambda$). While in solids, if a fixed-free boundary condition is applied, 1/8 wavelength corresponds exactly to the midpoint of a quarter-wavelength rod ($x_{opt}=\frac{1}{8}\lambda=\frac{1}{2}(\frac{1}{4}\lambda)=\frac{1}{2}L_{0.25}$). For a rod of $1.8m$ in length and $2.38mm$ in radius with a $2kg$ tip mass mounted at the end, the wavelength is approximately $\lambda=\frac{c}{f}\approx\frac{\sqrt{E/\rho}}{f}=\frac{5091}{92.8}\approx55m $, while $\lambda/8=6.86m$ is beyond the total length of the rod $L=1.8m$. Hence, in this case the optimal location of the stage approaches the end mass.

The thermal penetration thickness $\delta_k=\sqrt{\frac{2\kappa}{\omega \rho c_\epsilon}}$ indicates the distance, measured from the isothermal boundary, that heat can diffuse through. Solid particles that are outside this thermal layer do not experience radial temperature fluctuations and therefore do not contribute to building the instability. The value of the thermal penetration thickness $\delta_k$, or more specifically, the ratio of $\delta_k/R$ is a key parameter for the design of the system. Theoretically, the optimal value of this parameter is attained when the rod radius is equal to $\delta_k$. In fluids, good performance can be obtained for values of $2\delta_k$ to $3\delta_k$. Here below, we study the optimal value of this parameter for the two configurations above. 

In the quasi-1D case, once the material, the length of the rod, and the boundary conditions are selected, the frequencies of vibration of the rod (we are only interested in the frequency $\omega$ that corresponds to the mode selected to drive the thermoacoustic growth) is fixed. This statement is valid considering that the small frequency perturbation associated to the thermal oscillations is negligible. Under the above assumptions, also $\delta_k$ is fixed therefore the ratio $R/\delta_k$ can be effectively optimized by tuning $R$. Figure \ref{opt}b shows that a rod having $R=\frac{\delta_k}{0.56}\approx2\delta_k$ yields the highest growth ratio $\frac{\beta}{\omega}$ for both boundary conditions.
The above analysis shows that the optimal values of $x_k/L$ and $\delta_k/R$ are quantitatively equivalent to their counterparts in fluids.

Another important factor is the energy dissipation of the system. This is probably the element that differentiates more clearly the thermoacoustic process in the two media.
The mechanism of energy dissipation in solids, typically referred to as damping, is quite different from that occurring in fluids. Although in both media damping is a macroscopic manifestation of non-conservative particle interactions, in solids their effect can dominate the dynamic response. Considering that the thermoacoustic instability is driven by the first axial mode of vibration, some insight in the effect of damping in solids can be obtained by mapping the response of the rod to a classical viscously damped oscillator. The harmonic response of an underdamped oscillator is of the general form $x(t)=Ae^{i\Lambda_D t}$, where $i\Lambda_D$ is the system eigenvalue given by $i\Lambda_D=-\zeta\omega_0+i\sqrt{1-\zeta^2}\omega_0$, where $\omega_0$ is the undamped angular frequency, and $\zeta$ is the damping ratio. The damping contributes to the negative real part of the system eigenvalue, therefore effectively counteracting the thermoacoustic growth rate (which, as shown above, requires a positive real part). In order to obtain a net growth rate, the thermally induced growth (i.e. the thermoacoustic effect) must always exceed the decay produced by the material damping. Mathematically, this condition translates into the ratio $\frac{\beta}{\omega}>\zeta$.

For metals, the damping ratio $\zeta$ is generally very small (on the order of 1\% for aluminum \cite{Lazan}). By accounting for the damping term in the above simulations, we observe that the undamped growth ratio $\frac{\beta}{\omega}$ becomes one or two orders of magnitude lower than the damping ratio $\zeta$. Therefore, despite the relatively low intrinsic damping of the material the growth is effectively impeded.

Considering that dissipative forces exist also in fluids, then a logical question is why their effect is so relevant in solids to be able to prevent the thermoacoustic growth? Our analyses have highlighted two main contributing factors:
\begin{enumerate}
	\item In fluids, the dissipation is dominated by viscous losses localized near the boundaries. This means that while particles located close to the boundaries experience energy dissipation, those in the bulk can be practically considered loss-free. Under these conditions, even weak pressure oscillations in the bulk can be sustained and amplified. In solids, structural damping is independent of the spatial location of the particles (in fact it depends on the local strain). Therefore, the bulk can still experience large dissipation. In other terms, even considering an equivalent dissipation coefficient between the two media, the solid would always produce a higher energy dissipation per unit volume.
	\item The net work during a thermodynamic cycle in fluids is done by thermal expansion at high pressure (or stress, in the case of solids) and compression at low pressure \cite{Swift}. Thermal deformation in fluids and solids can occur on largely disparate spatial scales. This behavior mostly reflects the difference in the material parameters involved in the constitutive laws with particular regard to the Young's modulus and the thermal expansion coefficient. In general terms, a solid exhibits a lower sensitivity to thermal-induced deformations which ultimately limits the net work produced during each cycle, therefore directly affecting the growth rate of the system.
\end{enumerate}
In principle, we could act on both the above mentioned factors in order to get a strong thermoacoustic instability in solids. Nevertheless, damping is an inherent attribute of materials and it is more difficult to control. Therefore, unless we considered engineered materials able to offer highly controllable material properties, pursuing approaches targeted to reducing damping appears less promising. On the other hand, we choose to explore an approach that targets directly the net work produced during the cycle. 

\paragraph*{Multi-stage configuration.}
In the previous section, we indicated that thermoacoustics in solids is more sensitive to dissipative mechanisms because of the lower net work produced in one cycle.
In order to address directly this aspect, we conceived a multiple stage (here below referred to as \textit{multi-stage}) configuration targeted to increase the total work per cycle. As the name itself suggests, this approach simply uses a series of stages uniformly distributed along the rod. The separation distance between two consecutive stages must be small enough, compared to the fundamental wavelength of the standing mode, in order to not alter the phase lag between the temperature and velocity fields.

\begin{figure}[h]
	\centering
	\includegraphics[scale=0.45]{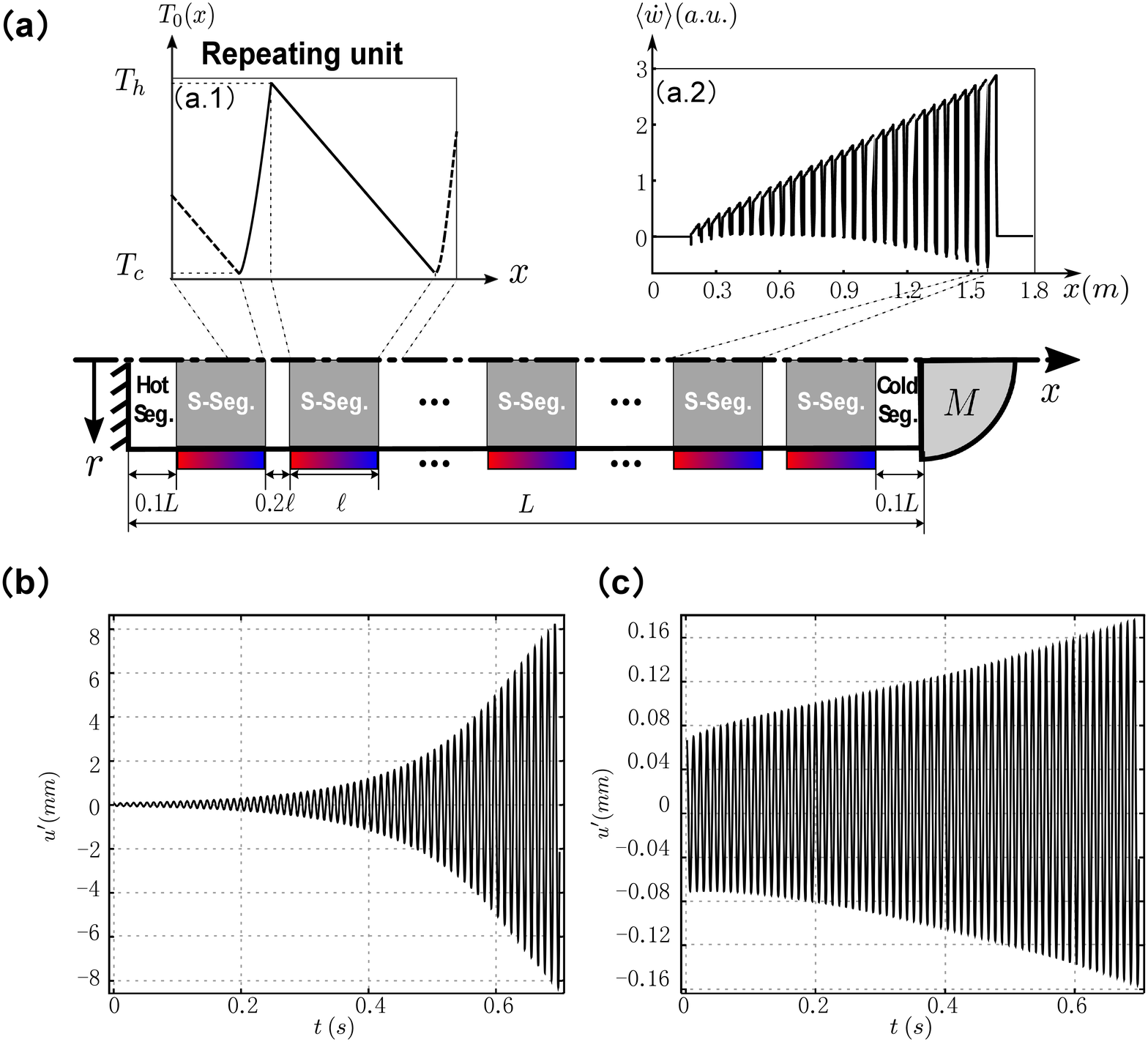}
	\caption{(\textbf{a}) Schematic diagram of the multi-stage configuration. The two insets show the mean temperature $T_0$ profile along the axial direction $x$ and the time averaged volume-change work $\langle{\dot{w}}\rangle$ (arbitrary scale and unit) along the rod. Time response at the moving end of a fixed-mass rod for the (\textbf{b}) undamped and {\bf(c)} $1\%$ damped configurations. }
	\label{multistack}
\end{figure}

We tested this design by numerical simulations using thirty stage elements located on the rod section $[0.1\sim0.9]$L, with $T_h=543.15K$ and $T_c=293.15K$ (Fig. \ref{multistack}a). The resulting mean temperature distribution $T_0(x)$ was a periodic sawtooth-like profile with a total temperature difference per stage $\Delta T=250K$. Note that, in the quasi-1D theory, in order to account for the finite length of each stage and for the corresponding axial heat transfer between the stage and the rod we tailored the gradient according to an exponential decay. In the full 3D numerical model, the exact heat transfer problem is taken into account with no assumptions on the form of the gradient. We anticipate that this gradient has no practical effect on the instability, therefore the assumption made in the quasi-1D theory has a minor relevance. A tip mass $M=0.353kg$ was used to reduce the resonance frequency and increase the wavelength so to minimize the effect of the discontinuities between the stages. 

The stability analysis performed according to the quasi-1D theory returned the fundamental eigenvalue as $i\Lambda_u=8.15+i598.6(rad/s)$ without considering damping, and $i\Lambda_d=2.27+i598.7(rad/s)$ with $1\%$ damping. Figure \ref{multistack}(a.2) shows the time averaged mechanical work $\langle{\dot{w}}\rangle$ along the rod. The elements in each stage do net work in each cycle. Although the segments between stages are reactive (because the non-uniform $T_0$ still perturbs the phase), their small size does not alter the overall trend. The positive growth rate obtained on the damped system shows that thermoacoustic oscillations can be successfully obtained in a damped solid if a multi-stage configuration is used.

Full 3D simulations were also performed to validate the multi-stage response. Figures \ref{multistack}b and \ref{multistack}c show the time response of the axial displacement fluctuation at the mass-end for both the undamped and the damped rods. The growth rates for the two cases are $\beta_u=6.87(rad/s)$ (undamped) and $\beta_d=1.28(rad/s)$ (damped). Contrarily to the single stage case, these results are in larger error with respect to those provided by the 1D solver. In the multi-stage configuration, the quasi-1D theory is still predictive but not as accurate. The reason for this discrepancy can be attributed to the effect of axial heat conduction. For the single stage configuration, the net axial heat flux $\kappa \frac{\partial^2 \hat{T}}{\partial x^2}$ is mostly negligible other than at the edges of the stage (see Fig. S4a). Neglecting this term in the 1D model does not result in an appreciable error. On the contrary, in a multi-stage configuration the existence of repeated interfaces where this term is non-negligible adds up to an appreciable effect (see Fig. S4b). This consideration can be further substantiated by comparing the numerical results for an undamped multi-stage rod produced by the 1D model and by the 3D model in which axial conductivity is artificially impeded. These two models return a growth ratio equal to $\beta_{1D} = 6.38(rad/s)$ and $\beta^{\kappa_x =0}_{3D} = 6.60(rad/s)$. More details are discussed in\cite{Suppl}.

\paragraph*{Conclusions.} In this study, we have theoretically and numerically shown the existence of thermoacoustic oscillations in solids. We presented a fully coupled, nonlinear, three-dimensional theory able to capture the occurrence of the instability and to provide deep insight into the underlying physical mechanism. The theory served as a starting point to develop a quasi-1D linearized model to perform stability analysis and characterize the effect of different design parameters, as well as a nonlinear 3D model. The occurrence of the thermoacoustic phenomenon was illustrated for a sample system consisting in a metal rod. Both models were used to simulate the response of the system and to quantify the instability. A multi-stage configuration was proposed in order to overcome the effect of structural damping, which is one of the main differences with respect to the thermoacoustics of fluids.

This study laid the theoretical foundation of thermoacoustics of solids and provided key insights into the underlying mechanisms leading to self-sustained oscillations in thermally-driven solid systems. It is envisioned that the physical phenomenon explored in this study could serve as the fundamental principle to develop a new generation of solid state thermoacoustic engines and refrigerators.


\bibliographystyle{Science}


\section*{Supplementary materials}
Materials and Methods\\
Supplementary Text\\
Figs. S1 to S5\\

\end{document}